\documentclass[aps,prl,twocolumn,preprintnumbers,amsmath,amssymb,superscriptaddress,showpacs,longbibliography]{revtex4-2}

\usepackage[colorlinks,bookmarks=true,citecolor=blue,linkcolor=red,urlcolor=blue,citecolor=blue]{hyperref}

\usepackage{graphicx}
\usepackage{subfigure}
\usepackage{amssymb,amsmath}
\usepackage{upgreek}
\usepackage{color,xcolor}
\usepackage{ulem}
\usepackage{lineno}
\usepackage{braket}

\begin{document}

\switchlinenumbers
\title{Dynamical Spectra of Spin Supersolid States in Triangular Antiferromagnets}

\author{Runze Chi}
\affiliation{Beijing National Laboratory for Condensed Matter Physics and Institute of Physics,
Chinese Academy of Sciences, Beijing 100190, China.}
\affiliation{School of Physical Sciences, University of Chinese Academy of Sciences, Beijing 100049, China.}

\author{Jiahang Hu}
\affiliation{Beijing National Laboratory for Condensed Matter Physics and Institute of Physics,
Chinese Academy of Sciences, Beijing 100190, China.}
\affiliation{School of Physical Sciences, University of Chinese Academy of Sciences, Beijing 100049, China.}

\author{Hai-Jun Liao}\email{navyphysics@iphy.ac.cn}
\affiliation{Beijing National Laboratory for Condensed Matter Physics and Institute of Physics,
Chinese Academy of Sciences, Beijing 100190, China.}
\affiliation{Songshan Lake Materials Laboratory, Dongguan, Guangdong 523808, China.}

\author{T. Xiang}\email{txiang@iphy.ac.cn}
\affiliation{Beijing National Laboratory for Condensed Matter Physics and Institute of Physics, Chinese Academy of Sciences, Beijing 100190, China.}
\affiliation{School of Physical Sciences, University of Chinese Academy of Sciences, Beijing 100049, China.}
\affiliation{Beijing Academy of Quantum Information Sciences, Beijing, 100190, China.}

\begin{abstract}
 We employ tensor network renormalization to explore the dynamical spectra of the easy-axis triangular-lattice antiferromagnet (TLAF) in a magnetic field. Our analysis identifies two distinct low-energy magnon excitations: a gapless Goldstone mode and a gapped mode. At zero field, the spectra display two nearly degenerate roton modes near the M point. With the increase of the magnetic field within the Y-shape superfluid phase, these modes diverge, with the roton excitation vanishing from the Goldstone mode branch, suggesting that the roton dip in this mode may just result from the energy-level repulsion imposed by the roton excitation in the gapped mode. Moreover, the in-plane spectral function shows substantial weight in high energies in the same spin excitation channel where the low-energy roton excitation appears. However, these roton excitations are absent in the V-shape supersolid phase.
\end{abstract}

\maketitle

\paragraph{Introduction---}
\noindent 
The quest to understand supersolidity, a remarkable state featuring both spatial order and superfluidity, has captivated researchers across disciplines. Originating from the study of solid helium-4~\cite{Leggett1970}, this enigmatic phenomenon has extended its reach to diverse systems, including ultracold gases~\cite{Boninsegni2012,Lu2015,Li2017,Julian2017,Tanzi2019,Norcia2021} and hard-core bosons~\cite{Wessel2005,Melko2005,Heidarian2005,Boninsegni2005,Wang2009}. A promising avenue for exploring supersolidity lies in the easy-axis S=1/2 triangular-lattice antiferromagnet (TLAF)~\cite{Heidarian2010,Sellmann2015,Gao2022}, where the spin states of magnetic ions mirror the occupancy states of lattice sites by Bose atoms. The ground state of the easy-axis TLAF, characterized by the spontaneous breaking of lattice translation and spin rotational symmetries, bears a resemblance to the supersolid state of Bose atoms. 

Recently, this elusive supersolid phase has been observed in the easy-axis triangular-lattice antiferromagnet Na$_2$BaCo(PO$_4$)$_2$, exhibiting a notable magnetocaloric effect~\cite{Xiang2024}.
Early thermodynamic measurements down to 50~mK at zero magnetic field~\cite{Zhong2019,Lee2021} suggested that Na$_2$BaCo(PO$_4$)$_2$ might be a candidate material for a quantum spin liquid (QSL). However, subsequent research~\cite{LiN2020,Sheng2022} revealed that at around 150~mK, Na$_2$BaCo(PO$_4$)$_2$ undergoes a transition from paramagnetic to antiferromagnetic state. Moreover, due to the small exchange interactions in Na$_2$BaCo(PO$_4$)$_2$, a moderate magnetic field can drive it into a fully polarized state. As a result, comprehensive inelastic neutron scattering measurements~\cite{Sheng2022,Sheng2024} have been conducted to explore its complete phase diagram under applied magnetic fields, including the Y-shape, up-up-down (UUD), V-shape, and fully polarized phases, where the Y-shape and V-shape phases possess supersolid order. The synthesis of other candidate materials for supersolid phases, such as K$_2$Co(SeO$_3$)$_2$~\cite{zhu2024,chen2024}, has also invigorated research in this field.

Despite these advancements, there remains a gap in understanding excitations of supersolid states in the easy-axis TLAF. While Na$_2$BaCo(PO$_4$)$_2$ exhibits clear long-range magnetic order at low-temperature~\cite{LiN2020,Sheng2022}, the recent neutron scattering experiments~\cite{Sheng2024} have revealed that, unlike the semiclassical spin wave theory predictions, its low-energy excitation spectrum lacks sharp magnon excitations and instead exhibits a broad continuum resembling deconfined fractional spinon excitations, suggesting the possibility of proximate quantum spin-liquid states~\cite{jia2023quantum}. Consequently, there is an urgent need for further theoretical exploration to determine the intrinsic nature of these spectral features and illuminate the elusive phenomenon of supersolidity.

To address these challenges, we conduct a comprehensive numerical investigation into the dynamical excitations of the easy-axis TLAF in a magnetic field by the tensor network method. We find that two sharp magnon excitation modes emerge at low energy: a gapless Goldstone mode and a gapped mode. Furthermore, two nearly degenerate roton or roton-like magnon excitation modes at M point are observed. As the magnetic field increases, the two nearly degenerate roton modes gradually separate, and the the roton mode from the lowest branch with Goldstone mode transitions from a dip to a hump shape, suggesting that the roton dip in this mode may just result from the energy-level repulsion imposed by the roton excitation in the second lowest branch. While another roton excitation persists throughout the entire Y-shape phase. Furthermore, the in-plane spectral function shows substantial high-energy spectral weight in the same spin excitation channel where this roton excitation appears, suggesting that the second branch of roton excitation stems from the magnon-Higgs scattering mechanism~\cite{Magnon-Higgs2015}. Finally, these roton excitations are absent in the V-shape supersolid phase.

\begin{figure}[t]
  \centering
  \includegraphics[width=0.48\textwidth]{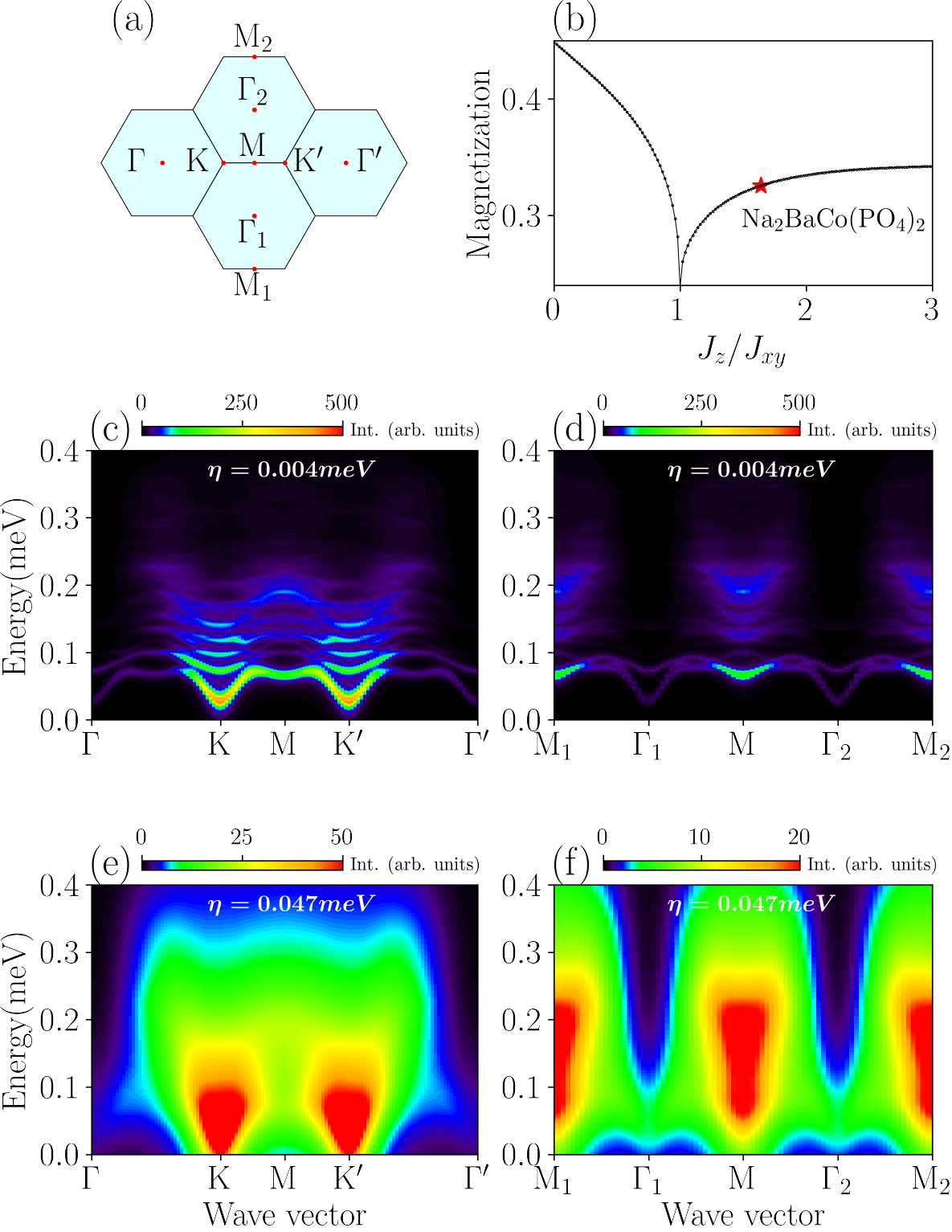}
  \caption{Zero-field spectral functions with two different Lorentzian broadening factors $\eta$ along two momentum paths. (a) Schematic diagram of the Brillouin zone. (b) Magnetization of LSWT as a function of anisotropic ratio $J_z/J_{xy}$, where the red star is corresponding to Na$_2$BaCo(PO$_4$)$_2$. 
  (c-d) The spectral functions with $\eta=0.004$ meV (about $0.053$ $J_{xy}$) along $\Gamma$-K-M-K$^{\prime}$-$\Gamma^{\prime}$ and M$_1$-$\Gamma_1$-M-$\Gamma_2$-M$_2$, respectively.
  (e-f) The spectral functions along two directions with $\eta=0.047$ meV (about $0.62$ $J_{xy}$), which is equal to the experimental energy resolution~\cite{Sheng2024}. 
\label{fig1}}
\end{figure}

\begin{figure*}[htbp]
  \centering
  \includegraphics[width=0.85\textwidth]{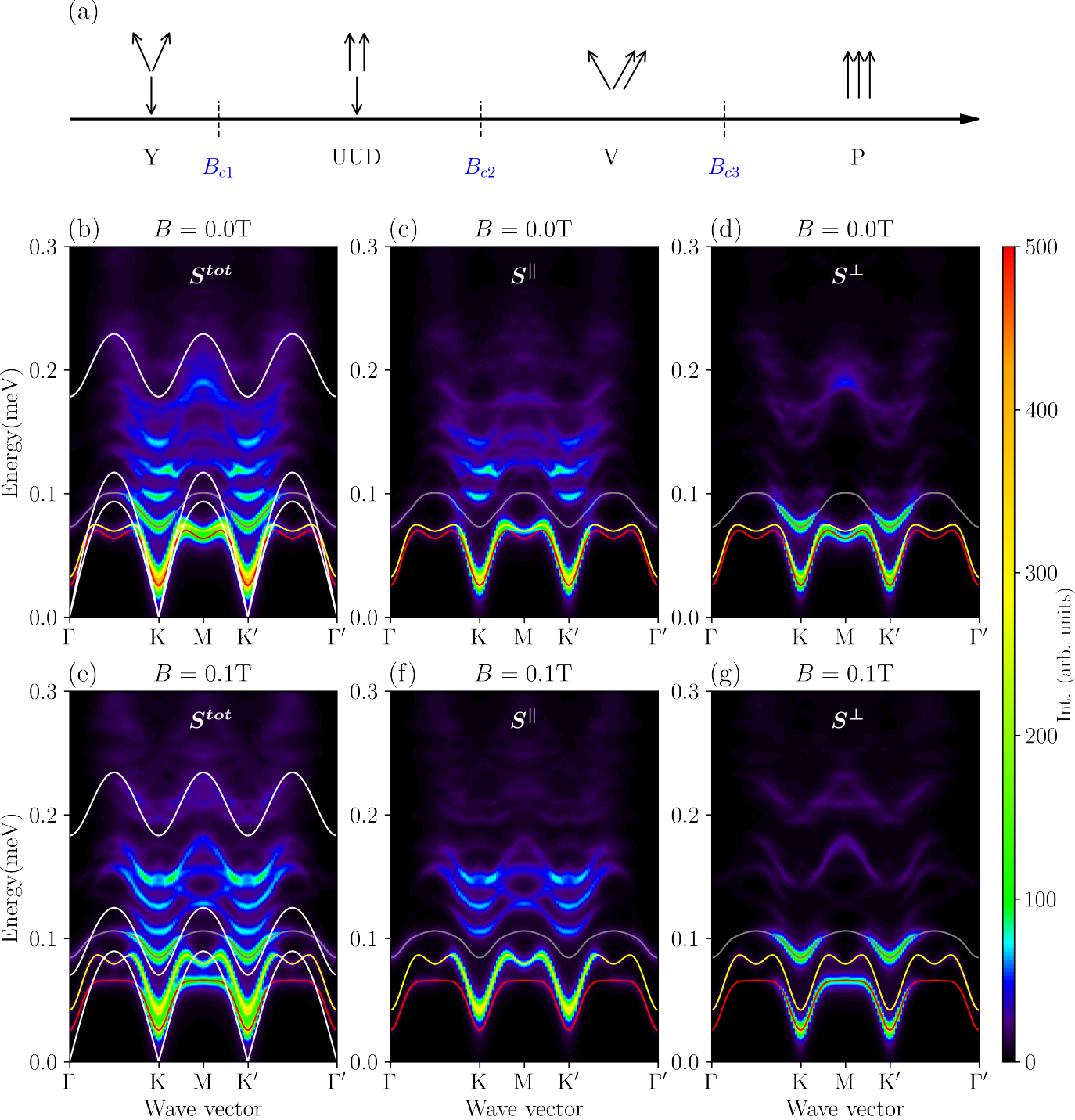}
  \caption{
    The dynamical spectral functions with $\eta=0.004$ meV in the Y-shape supersolid phase. (a) The phase diagram of the easy-axis TLAF. For Na$_2$BaCo(PO$_4$)$_2$ with $J_{xy}=0.076$~meV, $J_{z}=0.125$~meV and $g_z=4.645$, the three critical points are $B_{c1}=0.42$ T, $B_{c2}=1.13$ T, and $B_{c3}=1.8$ T, respectively. (b)-(d) The spectral function at zero field. (e)-(g) The spectal function with $B=0.1$ T. (b) and (e) show the total spectral functions $S^{tot}(\boldsymbol{k},\omega)$, where the white lines denote the LSWT results, and the red, yellow, and gray lines correspond to the lowest-energy three branches of magnon excitations obtained by tensor network method, respectively. (c) and (f) show the in-plane spectral function $S^{\parallel}(\boldsymbol{k},\omega)=S^{xx}(\boldsymbol{k},\omega)+S^{zz}(\boldsymbol{k},\omega)$. (d) and (g) are the out-of-plane spectral function $S^{\perp}(\boldsymbol{k},\omega)=S^{yy}(\boldsymbol{k},\omega)$, which probes the Goldstone mode.
  \label{fig2}}
  \end{figure*}

\begin{figure*}[htbp]
  \centering
  \includegraphics[width=0.85\textwidth]{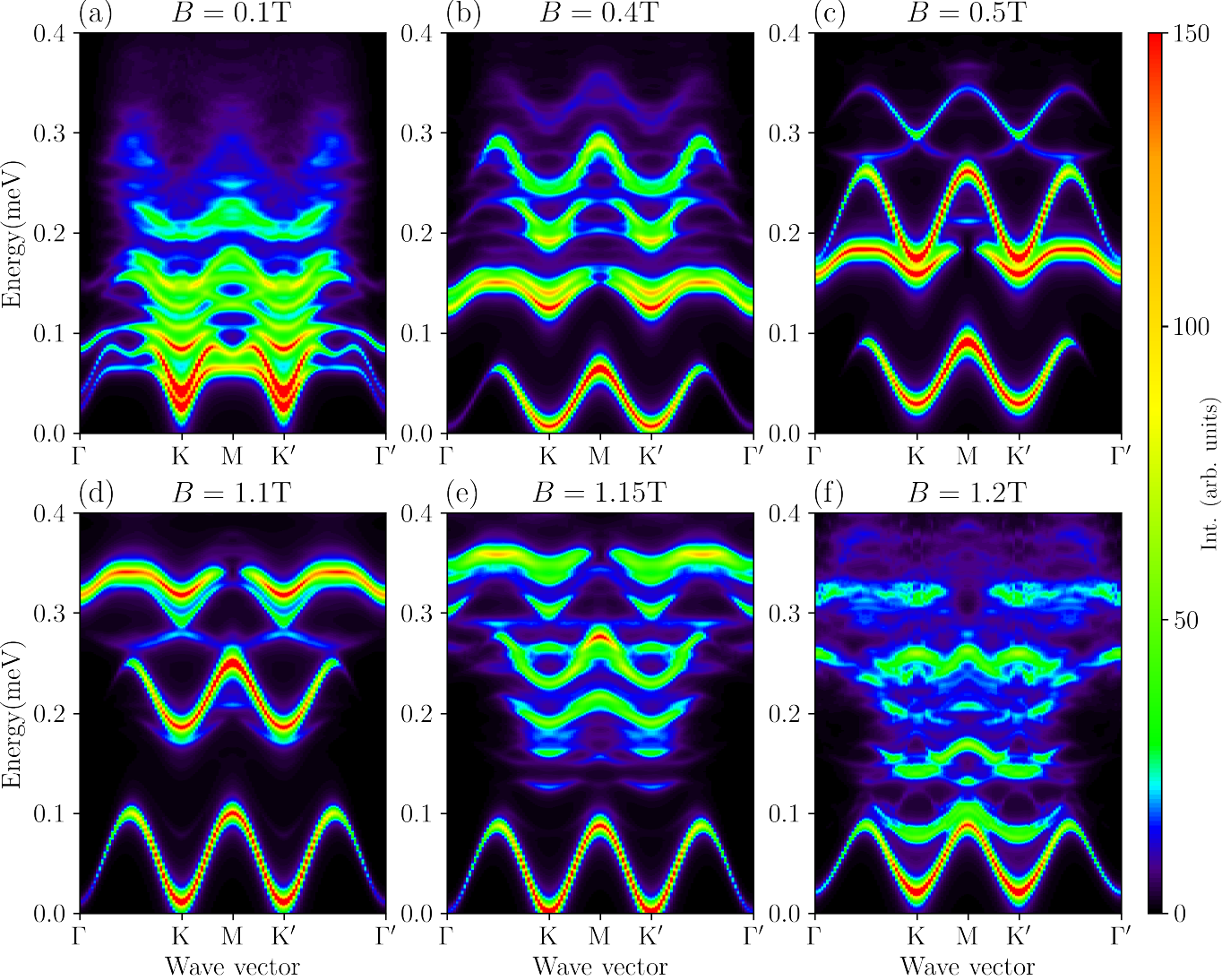}
  \caption{Comparison of the spectral functions with $\eta=0.004$ meV between the two supersolid phases and UUD phase. (a)-(b) The spectral functions in the Y-shape supersolid phase at magnetic fields $B=0.1$ T and $B=0.4$ T, respectively. (c)-(d) The spectral functions in the UUD phase with magnetic fields $B=0.5$ T and $B=1.1$ T, respectively. (e)-(f) The spectral functions in the V-shape supersolid phase with magnetic fields $B=1.15$ T and $B=1.2$ T, respectively.
  \label{fig3}}
  \end{figure*}

\bigskip
\paragraph{Model and method---}
\noindent
We here consider the easy-axis triangular-lattice antiferromagnetic XXZ model
\begin{equation}
\hspace{-2mm}\mathcal{H} \!=\! \sum_{\langle i j\rangle}[J_{z} S_{i}^{z} S_{j}^{z} + J_{xy} \left(S_{i}^{x} S_{j}^{x}\!+\!S_{i}^{y} S_{j}^{y}\right)] \!-\! g_z \mu_B B \sum_i S_{i}^{z},
\label{Ham}
\end{equation}
 where $\langle i j \rangle$ runs over all the nearest-neighbor sites of the two-dimensional triangular lattice, and $J_z > J_{xy}$. The ground-state phase diagram of this model has been studied~\cite{Yamamoto2014,Sellmann2015}. As the magnetic field increases, as shown in Fig.~\ref{fig2}(a), the ground state of this model undergoes transitions through the Y-shape phase, up-up-down (UUD) phase, V-shape phase, and fully polarized phase. Among them, the Y- and V-shape phases are spin supersolid phases~\cite{Gao2022,Xiang2024}. In this work, we adopt typical anisotropic parameters $J_{xy}=0.076$ meV, $J_{z}=0.125$ meV and $g_z = 4.645$, which are believed to precisely describe the supersolid material of the easy-axis triangular antiferromagnet Na$_2$BaCo(PO$_4$)$_2$~\cite{Sheng2024}. Although the following discussion focuses on a specific material, the outcomes are applicable to other easy-axis triangular antiferromagnets that have the same phase diagram.
 
To simulate the spin excitation spectra of neutron scattering experiment, we utilize the state-of-art tensor-network method~\cite{Vanderstraeten2015,Vanderstraeten2019,Ponsioen2020,Ponsioen2022,Chi2022}, based on the single-mode excited tensor network representation~\cite{Ostlund1995} and automatic differentiation~\cite{Liao2019}, to calculate the spin spectral function
\begin{eqnarray}
&&  S^{tot}(\boldsymbol k , \omega ) = \sum_\alpha S^{\alpha\alpha} (\boldsymbol k, \omega) , \quad (\alpha = x, y, z) , \nonumber \\
&&  S^{\alpha\beta} (\boldsymbol k, \omega) = \langle 0 | S_{-\boldsymbol k}^{\alpha} \delta(\omega-H+E_0)S_{\boldsymbol k}^{\beta} |0\rangle. 
\end{eqnarray}
 This method has recently been successfully applied to the easy-plane triangular-lattice antiferromagnet Ba$_3$CoSb$_2$O$_9$~\cite{Chi2022}, and quantitatively accounts for its neutron scattering experimental results. Further details on this method can be found in the Supplemental Material (SM) and Refs.~\cite{Ponsioen2022,Chi2022}.  
 
 In the calculation, the delta function is expanded by a Lorentzian broadening factor $\eta$, which can mimic the broadening effect by the instrument energy resolution in neutron scattering experiment. It is crucial to distinguish whether the low-energy continuum spectrum is intrinsic or caused by finite instrumental resolution.

\paragraph{Results---}
\noindent
Figure~\ref{fig1} shows the zero-field spin excitation spectra with different Lorentzian broadening factor $\eta$ along $\Gamma$-K-M-K$^{\prime}$-$\Gamma^{\prime}$ and M$_1$-$\Gamma_1$-M-$\Gamma_2$-M$_2$.
As shown in Fig.~\ref{fig1}(c) and (d), sharp magnon excitations are observed along both directions in the low-energy excitation spectra, consistent with the conventional expectations for the low-energy excitations of magnetically ordered systems but inconsistent with the low-energy continuum spectra observed in recent neutron scattering experiments~\cite{Sheng2024}. To account for this discrepancy, we investigated the excitation spectra using a broadening factor of $0.047$ meV (about 0.62 $J_{xy}$), corresponding to the energy resolution of recent neutron scattering experiment~\cite{Sheng2024}. Under this condition, as shown in Fig.~\ref{fig1}(e) and (f), the spin excitation spectra exhibit obvious low-energy continuum, similar to the experimental results. Besides, we analyzed the variation of magnetization with anisotropy and found that regardless of whether it's easy-plane or easy-axis anisotropy, the magnetization is always enhanced as it deviates from the isotropic Heisenberg point (see Fig.~\ref{fig1}(b)). This indicates that magnetic anisotropies tend to stabilize magnetic order rather than suppress it, suggesting that Na$_2$BaCo(PO$_4$)$_2$ is further away from the U(1) Dirac spin liquid phase compared to the Heisenberg point. Therefore, these results suggest that the low-energy continuum spectra observed in the experiment are not intrinsic but likely caused by broadening effects due to instrumental resolution or other factors.

Furthermore, we find that these two low-energy magnon excitations can be distinguished by different components of spectral function. As shown in Fig.~\ref{fig2} (d) and (g), the lowest-energy magnon excitation (red curve) predominantly appears in the out-of-plane spectral function
\begin{equation}
S^{\perp}(\boldsymbol{k},\omega) = S^{yy}(\boldsymbol{k},\omega),
\end{equation}
while the next lowest-energy magnon excitation (yellow curve) mainly appears in the in-plane spectral function (see Fig.~\ref{fig2} (c) and (f))
\begin{equation}
S^{\parallel}(\boldsymbol{k},\omega) = S^{xx}(\boldsymbol{k},\omega) + S^{zz}(\boldsymbol{k},\omega).  
\end{equation}
The out-of-plane spectral function represents transverse fluctuations perpendicular to the xz-plane of magnetic order. Therefore, it should exhibit a gapless Goldstone mode at the K point. The reason why the lowest-energy branch (red curve) shows a finite gap at the K point is that the tensor network method introduces a truncation parameter $D$, which leads to a finite correlation length of the ground state, resulting in a finite excitation gap. Practical computations~\cite{Vanderstraeten2019,Ponsioen2020,Ponsioen2022,Chi2022,Tu2024,Wang2024} have demonstrated that this gap induced by finite $D$ systematically decreases and gradually vanishes with increasing $D$ (see Fig.~S2 in the Sec. II of SM), thereby satisfying the Goldstone theorem. As shown by the white curves in Fig.~\ref{fig2}(b), LSWT predicts that there are two gapless magnon excitations around K point~\cite{LSWT1992}, where the first branch is the well-known Goldstone mode due to breaking the U(1) symmetry of the Hamiltonian Eq.~(\ref{Ham}), and the second gapless mode is because the classical ground state possesses an additional U(1) rotational symmetry around the y-axis (where the spins are aligned in the xz-plane)~\cite{LSWT1992}. The second gapless mode only exists within the linear spin-wave approximation. In the presence of quantum fluctuation or magnetic field, as shown in Fig.~\ref{fig2}(c) and (f), the ground state no longer exhibits additional U(1) rotational symmetry around the y-axis, and thus the second-lowest branch acquires a finite gap, which is about $0.01$ meV at zero field (see Fig.~S2 of SM).

 More importantly, the low-energy excitation spectrum at zero field exhibits two nearly degenerate branches of roton excitation modes (an elementary excitation first seen in superfluid helium-4) with a minimum at M point. It can be seen from Fig.~\ref{fig2}(c-d) that these two nearly degenerate roton excitation modes are one-to-one connected to the lowest-energy Goldstone mode and the second lowest-energy gapped mode at K point. Besides superfluid helium-4, the roton excitation has also been observed in the triangular antiferromagnet Ba$_3$CoSb$_2$O$_9$~\cite{Ito2017,Macdougal2020} and square-lattice antiferromagnet~\cite{Piazza2015}. There have been many theoretical attempts to explain the origin of roton excitation, such as the fractional spinon theory~\cite{Zheng2006,Zhang2020,Ferrari2019,Ghioldi2022}, vortex–antivortex pair~\cite{LiHan2020, Alicea2006}, avoid quasiparticle decay~\cite{Verresen2019} and magnon-Higgs scattering~\cite{Magnon-Higgs2015,Chernyshev2009, Chi2022}. As shown in Fig.~\ref{fig2} (d), the roton excitation in the branch of Goldstone mode (red curve) stems from the out-of-plane fluctuation $S^{\perp}(\boldsymbol{k},\omega)$. It quickly separates from the second lowest branch and disappears with the increasing of magnetic field (see Fig.~S4 (k-o)), suggesting that it arises from the energy-level repulsion induced by the roton excitation in the second lowest branch. While another roton excitation in the second lowest branch (yellow curve) arises from the in-plane fluctuation $S^{\parallel}(\boldsymbol{k},\omega)$ (see Fig.~\ref{fig2} (c)), and it still persists in the whole Y phase (see Fig.~S4 (f-j)). Besides, the in-plane spectral function shows substantial high-energy spectral weight in the same spin excitation channel where this roton excitation appears, suggesting that the emergence of the this roton excitation is more likely attributable to the strong magnon-Higgs scattering mechanism~\cite{Magnon-Higgs2015} between the in-plane transverse magnon mode and the in-plane longitudinal Higgs mode.

To further demonstrate the existence of strong magnetic-Higgs scattering in the supersolid phase, we compare the excitation spectra between two supersolid phases and the UUD phase near their phase boundaries. In the UUD phase, since all spins of the ground state are aligned collinearly, there are no three-magnon interaction terms in the spin wave expansion, resulting in weak coupling between transverse and longitudinal fluctuation modes and, thus, no prominent continuum spectra (see Fig.~\ref{fig3} (c) and (d)). However, once entering the supersolid phases, the high-energy excitation spectra in the two supersolid phases begin to become more diffusive. The further away from the UUD phase, the more severe the diffusion becomes (see Fig.~\ref{fig3} (a-b) and (e-f)). This is because the spins of the ground state are non-collinearly arranged in both the Y-shape and V-shape supersolid phases, the three-magnon interaction term inevitably appears in the spin-wave expansion. This leads to strong coupling between transverse magnon mode and longitudinal Higgs mode, thereby resulting in continuum spectra. In addition, we also note that the low-energy excitation spectra in the V-shape supersolid phase do not exhibit roton excitations, indicating that roton excitations are not a necessary condition for the onset of supersolid order, unlike in traditional superfluid helium-4.

\bigskip
\paragraph{Discussion---}
\noindent
This work presents a comprehensive numerical investigation into the dynamical excitations of the easy-axis triangular-lattice antiferromagnets under a magnetic field, employing the tensor network method. The study reveals that sharp magnon excitations emerge at low energies, indicating that the observed low-energy continuum spectra likely stem from broadening effects of the instrumental energy resolution, especially when exchange interaction is weak. It reminds us to be very cautious when a magnetic material with relatively weak exchange interactions exhibits low-energy continuum spectra in neutron scattering experiments. The low-energy continuum may not necessarily be a signature of fractionalized spinon excitations, but could also be due to broadening effects caused by other factors.

 Moreover, we find two branches of roton excitations at zero field. The roton excitation in the branch of Goldstone mode is rapidly destroyed by the magnetic field. Conversely, the roton excitation arising from in-plane fluctuations remains stable throughout the entire Y-shape phase under the influence of the magnetic field. Further insights from the in-plane spectral function reveal a close connection between the roton excitations and high-energy continuum spectra, suggesting that the roton excitations are caused by the mechanism of strong magnon-Higgs scattering~\cite{Magnon-Higgs2015}. On the other hand, the absence of roton-like excitations in the V-shape supersolid phase suggests that it is not inherently linked to the spin supersolid phase. These findings not only enhance our understanding of spin dynamics in TLAFs but also provide new insights into the underlying mechanism of supersolid states, paving the way for potential applications that leverage the novel properties of supersolid states.

\bigskip
\paragraph{\bf Acknowledgements}
\noindent
We thank Yuan Wan and Wei Li for the helpful discussion.
This work is supported by the Strategic Priority Research Program of Chinese Academy of Sciences (Grants No.~XDB0500202, ~XDB33010100 and No.~XDB33020300), the National Natural Science Foundation of China (Grants No.~12322403, No.~12347107, No.~12488201, No.~11874095, and No.~11974396), the National Key Research and Development Project of China (Grants No.~2022YFA1403900 and No.~2017YFA0302901) and the Youth Innovation Promotion Association of Chinese Academy of Sciences (Grant No.~2021004).

\bibliographystyle{apsrev4-2}
\bibliography{Supersolid}

\clearpage

\addtolength{\oddsidemargin}{-0.75in}
\addtolength{\evensidemargin}{-0.75in}
\addtolength{\topmargin}{-0.725in}

\newcommand{\addpage}[1] {
 \begin{figure*}
   \includegraphics[width=8.5in,page=#1]{Supersolid_SM.pdf}
 \end{figure*}
}
\addpage{1}
\addpage{2}
\addpage{3}
\addpage{4}

\end{document}